\documentclass[twocolumn,prb,color,superscriptaddress,psfig,showpacs,amsmath,amssymb,nobibnotes]{revtex4-1}
\setcitestyle{square,numbers}
\usepackage{graphicx}% Include figure files
\usepackage{float}
\usepackage{epstopdf}
\usepackage{dcolumn}% Align table columns on decimal point
\usepackage{hyperref}% add hypertext capabilities; optional: [hidelinks]
\hypersetup{
    colorlinks   = true, %Colours links instead of boxes
    urlcolor     = blue, %Colour for external hyperlinks
    linkcolor    = blue, %Colour of internal links
    citecolor    = blue %Colour of citations
}
\usepackage{color}
\usepackage{amsmath}
\usepackage{xspace}
\newcommand{\DMTTF}{$o$\,-DMTTF\xspace}
\newcommand{\DMTTFX}{($o$\,-DMTTF)$_2X$\xspace}
\newcommand{\DMTTFCl}{($o$\,-DMTTF)$_2$Cl\xspace}
\newcommand{\DMTTFBr}{($o$\,-DMTTF)$_2$Br\xspace}
\newcommand{\TSP}{$T_{\text{SP}}$\xspace}
\newcommand{\etal}{\textit{et al.}\xspace}

\begin{document}
	
\title{Coherent spin dynamics of solitons in the organic spin chain compounds ($o$-DMTTF)$_2X$ ($X$ = Cl, Br)}

\author{J.~Zeisner}
\altaffiliation[]{j.zeisner@ifw-dresden.de}
\affiliation{Leibniz Institute for Solid State and Materials Research IFW Dresden, D-01069 Dresden, Germany}
\affiliation{Institute for Solid State and Materials Physics, TU Dresden, D-01062 Dresden, Germany}
\author{O.~Pilone}
\affiliation{Aix-Marseille Universit\'{e}, CNRS, IM2NP UMR 7334, F-13397 Marseille, France}
\author{L.~Soriano}
\affiliation{Aix-Marseille Universit\'{e}, CNRS, IM2NP UMR 7334, F-13397 Marseille, France}
\author{G.~Gerbaud}
\affiliation{Aix-Marseille Universit\'{e}, CNRS, BIP UMR 7281, F-13402 Marseille, France}
\author{H.~Vezin}
\affiliation{Universit\'{e} de Lille, CNRS, LASIR UMR 8516, F-59655 Villeneuve d'Ascq, France}
\author{O.~Jeannin}
\affiliation{Universit\'{e} de Rennes, CNRS, ISCR UMR 6226, F-35042 Rennes, France}
\author{M.~Fourmigu\'{e}}
\affiliation{Universit\'{e} de Rennes, CNRS, ISCR UMR 6226, F-35042 Rennes, France}
\author{B.~B\"{u}chner}
\affiliation{Leibniz Institute for Solid State and Materials Research IFW Dresden, D-01069 Dresden, Germany}
\affiliation{Institute for Solid State and Materials Physics, TU Dresden, D-01062 Dresden, Germany}
\author{V.~Kataev}
\altaffiliation[]{v.kataev@ifw-dresden.de}
\affiliation{Leibniz Institute for Solid State and Materials Research IFW Dresden, D-01069 Dresden, Germany}
\author{S.~Bertaina}
\altaffiliation[]{sylvain.bertaina@im2np.fr}
\affiliation{Aix-Marseille Universit\'{e}, CNRS, IM2NP UMR 7334, F-13397 Marseille, France}
\date{\today}

\begin{abstract}
We studied the magnetic properties, in particular dynamics, of the correlated spins associated with natural defects in the organic spin chain compounds ($o$\,-DMTTF)$_2X$ ($X =$ Br, Cl) by means of electron spin resonance (ESR) spectroscopy. Both materials exhibit spin-Peierls transitions at temperatures around 50\,K \mbox{[P. Foury-Leylekian \textit{et al.}, \href{http://dx.doi.org/10.1103/PhysRevB.84.195134}{Phys. Rev. B \textbf{84}, 195134 (2011)}]}, which allow a separation of the properties of defects inside the chains from the magnetic response of the spin chains. Indeed, continuous wave ESR measurements performed over a wide temperature range evidence the evolution of the spin dynamics from being governed by the spins in the chains at elevated temperatures to a low-temperature regime which is dominated by defects within the spin-dimerized chains. Such defects polarize the antiferromagnetically coupled spins in their vicinity, thereby leading to a finite local alternating magnetization around the defect site which can be described in terms of a soliton, i.e. a spin 1/2 quasiparticle built of many correlated spins, pinned to the defect. In addition, contributions of triplon excitations of the spin-dimerized state to the ESR response below the transition temperature were observed which provides a spectroscopic estimate for the spin-gap of the studied systems. Moreover, details of spin dynamics deep in the spin-Peierls phase were investigated by pulse ESR experiments which revealed Rabi-oscillations as signatures of coherent spin dynamics. The latter is a prerequisite for a selective manipulation of the defect-induced soliton spin states which is, for instance, relevant in the context of quantum computation. From a comparison of the characteristic damping times of the Rabi oscillations with measurements of the spin relaxation times by means of primary-echo decay and CPMG methods it becomes evident that inhomogeneities in local magnetic fields strongly contribute to the soliton decoherence. 
\end{abstract}

\maketitle

\section{Introduction}
\begin{figure*}[t]
	\centering
	\includegraphics*[width=0.97\textwidth]{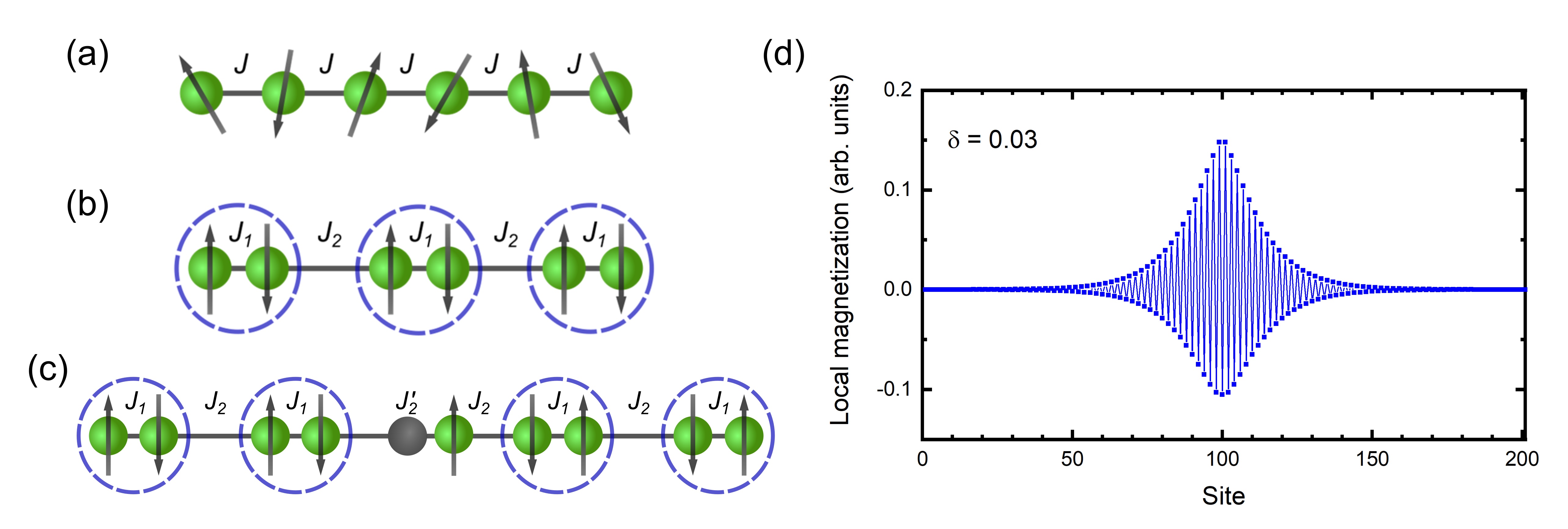}
	\caption{Schematic representation of the spin-Peierls transition and defect-induced soliton formation. (a) At temperatures above the spin-Peierls temperature \TSP the uniform Heisenberg chain consists of equidistant spins coupled by an isotropic nearest-neighbor exchange $J$. (b) For $T <$ \TSP the bond lengths within the chain lattice are modulated, eventually leading to the formation of spin dimers (indicated by blue circles). The antiferromagnetic intradimer interaction $J_1 = J(1+\delta)$ is larger than the interdimer interaction $J_2 = J(1-\delta)$ giving rise to the spin-singlet ground state of the dimerized chain. \mbox{(c) A nonmagnetic} defect (gray sphere) introduced into a dimerized spin system breaks the chains into finite segments and can be modeled, for instance, by two successive weak bonds $J_2'$ and $J_2$ (with $J_2' < J_2$) within the chain. The coupling constant $J_2'$ describes the effective exchange coupling between two spins across the defect site. (d) As a result of such a defect, a finite local alternating magnetization is induced around the defect site. This can be described in terms of a quasiparticle, the pinned soliton, which involves many entangled spins in the vicinity of the defect but carries an overall spin of 1/2. The profile of the local alternating magnetization shown here was calculated using a density matrix renormalization group method (see Ref.~\cite{Bauer2011} for further details) considering a defect at the center of a dimerized chain with a length of 201 sites and assuming $J_2' = J_2$ as well as a dimerization parameter $\delta = 0.03$.}
	\label{fig:spin-Peierls_chain}
\end{figure*}

Defects in one-dimensional spin systems continue to be an active field of modern solid state research as they are able to alter magnetic properties of the hosting materials drastically \cite{Alloul2009}. For example, nonmagnetic defects introduced into isotropic antiferromagnetic (AFM) spin chains lead to a breaking of the chains into finite segments and thereby induce additional paramagnetic-like moments in the spin systems (see, e.g., Refs.~\cite{Alloul2009,Sirker2007}). In spin-Peierls systems which feature a nonmagnetic ground state due to a dimerization of the (spin) lattice \cite{Bray1975} similar effects may appear, as illustrated in Fig.~\ref{fig:spin-Peierls_chain}. In addition, such a nonmagnetic singlet ground state enables a detailed study of the defect-induced magnetic properties as the influence of the spins residing on the regular sites of the chains is strongly reduced. Defects in the spin-Peierls chains polarize the antiferromagnetically coupled spins in their vicinity, thereby leading to a finite local alternating magnetization around the defect site \cite{Khomskii1996,Nishino2000} [see Figs.~\ref{fig:spin-Peierls_chain}(c) and (d)]. The latter can be described in terms of a soliton, i.e. a spin 1/2 quasiparticle built of many correlated spins, pinned to the defect or chain break \cite{Khomskii1996,Nishino2000}. As a consequence of the defect-induced soliton formation, a spin-Peierls system can show a magnetic response at low temperatures despite the nonmagnetic character of the ground state and of the introduced impurity. This paramagnetic-like behavior of the soliton can be evidenced, for instance, by temperature-dependent susceptibility measurements at low temperatures. In contrast to isolated, non-interacting magnetic impurities in a three-dimensional lattice which are frequently observed in real materials, a soliton is of a purely quantum mechanical nature. Moreover, since the solitons are formed from entanglement of many spins within the dimerized chain, their properties, in particular their extension along the chain sites as well as the coupling strength between adjacent solitons, are intimately connected to the characteristics of the chain, i.e., the coupling $J$ and the dimerization parameter $\delta$ \cite{Khomskii1996,Nakano1980,Dobry1997}. Theoretically, an isotropic one-dimensional spin system that shows a spin-Peierls transition at low temperatures could be modeled in terms an $S = 1/2$ alternating-exchange Heisenberg chain Hamiltonian \cite{Barnes1999,Johnston2000}
\begin{equation}
\label{eq:HAFM-Hamiltonian}
\mathcal{H} = \sum_{i}J(1+\delta)\boldsymbol{S}_{2i-1}\cdot \boldsymbol{S}_{2i} + J(1-\delta)\boldsymbol{S}_{2i}\cdot\boldsymbol{S}_{2i+1} \ \ \ .
\end{equation}
Throughout this work we consider an isotropic Heisenberg type exchange coupling $J$ with $J > 0$ corresponding to AFM coupling between nearest neighbor spins $\boldsymbol{S}_i$ and $\boldsymbol{S}_j$. The dimerization parameter of the chain is defined such that $\delta = 0 $ recovers the Heisenberg chain above the spin-Peierls transition temperature \TSP [Fig.~\ref{fig:spin-Peierls_chain}(a)] while the limit $\delta = 1$ corresponds to isolated dimers with a singlet ground state [Fig.~\ref{fig:spin-Peierls_chain}(b)]. For real spin-Peierls \mbox{systems}, $\delta$ is expected to be in between these two limits. Note that $J$ and $\delta$ may show a pronounced temperature dependence below \TSP since the magnetic transition is caused by a deformation of the crystal lattice, i.e., due to spin-phonon interactions (see, e.g., Ref.~\cite{Johnston2000}). These two parameters are, in principle, accessible via a quantitative analysis of the temperature-dependent spin susceptibility if the system can be appropriately described by the Hamiltonian in Eq.~(\ref{eq:HAFM-Hamiltonian}) \cite{Johnston2000}. Thus, temperature-dependent measurements of the static susceptibility or the electron spin resonance (ESR) intensity provide a means to characterize the static magnetic properties of solitons in real spin-Peierls materials. The latter are usually organic compounds (see, for instance, Refs.~\cite{Pouget2012,KrugvonNidda2009}), the only well-established inorganic spin-Peierls material up to now being CuGeO$_3$ (see, e.g., Ref.~\cite{Uchinokura2002}). Previous studies of the magnetic properties of the organic salts (TMTTF)$_2X$ ($X =$ AsF$_6$, PF$_6$) \cite{Bertaina2014} indeed could assign the magnetic low-temperature response to the formation of solitons pinned to defect sites which are naturally present in these crystals. Moreover, a remarkable soliton spin dynamics was revealed featuring a substantial spin coherence as evidenced by the observation of Rabi oscillations in pulse ESR measurements \cite{Bertaina2014}. Such a coherent dynamics is a prerequisite for a selective manipulation of the soliton spin states. The robustness of the Rabi oscillations with respect to the presence of anisotropic dipolar interactions, which usually result in decoherence of the spins (see, e.g., Ref.~\cite{DeRaedt2012}), was attributed to the existence of isotropic exchange interactions between the solitons which, in turn, result from the fact that the solitons are built of an ensemble of strongly correlated spins \cite{Bertaina2014}. This effect renders the spin dynamics of the solitons observed in (TMTTF)$_2$AsF$_6$ and (TMTTF)$_2$PF$_6$ comparable to the one encountered in inorganic systems of magnetic ions diluted in nonmagnetic matrices \cite{Bertaina2007,Nellutla2007,Rakhmatullin2009,DeRaedt2012,Baibekov2011,Baibekov2017} which were proposed as realizations of qubits with potential applications in the field of quantum computation \cite{Bertaina2007}. In this respect, as well as for the sake of deepening the understanding of spin dynamics of pinned solitons in dimerized spin chains, it is important to verify whether the existence of a coherent soliton spin dynamics represents a generic property of organic spin chain systems with natural defects. In the present work, we address this question by ESR studies on \DMTTFCl and \DMTTFBr \cite{Fourmigue2008}, two members of a related family of one-dimensional organic spin-Peierls compounds (\mbox{$o$-DMTTF} stands for the \textit{ortho}-dimethyltetrathiafulvalene molecule).

This paper is organized as follows: Details on the crystal structure of the title compounds as well as on experimental methods are provided in Sec.~\ref{sec:methods}. The results of our experimental investigations are presented and discussed in Sec.~\ref{sec:results} starting with the magnetic characterization of the spin systems by means of temperature- and angular-dependent continuous wave (CW) ESR measurements in Sec.~\ref{subsec:results_CW_ESR}. This is followed by the results of pulse ESR experiments in Sec.~\ref{subsec:results_pulse_ESR} which reveal the coherent character of the spin dynamics in \DMTTFX. Main conclusions of the present work are summarized in Sec.~\ref{sec:Conclusions} while a discussion of the temperature-dependent spin susceptibility and their analysis is provided in the Appendix.

\section{Samples and Methods}
\label{sec:methods}

\begin{figure}
	\centering
	\includegraphics[width=0.9\columnwidth]{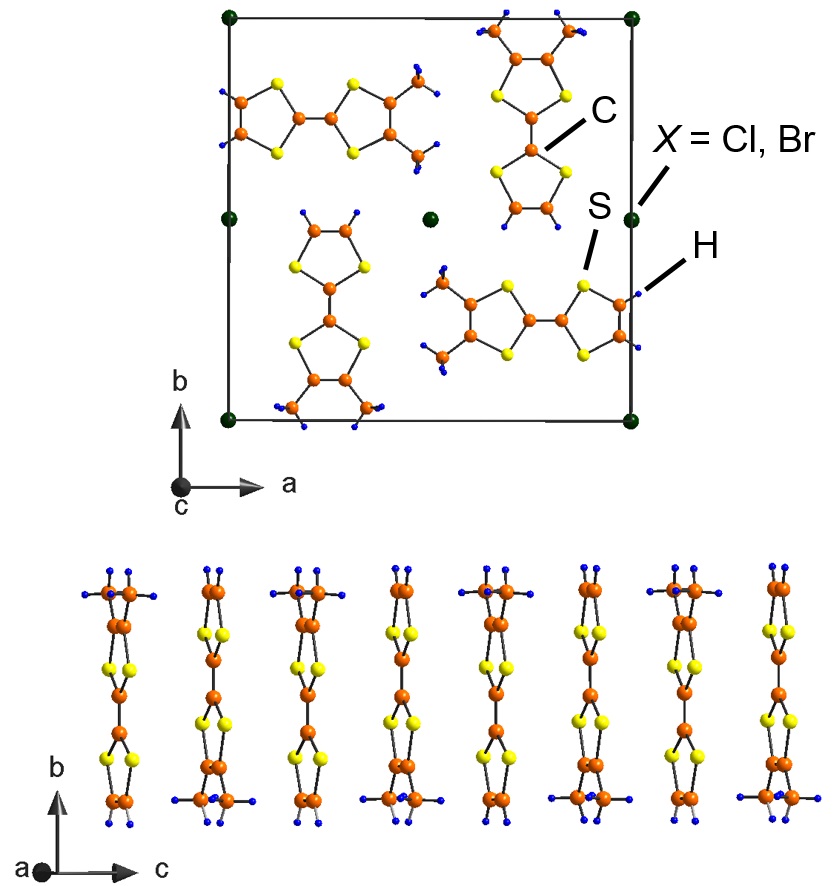}
	\caption{Crystallographic structure of the \DMTTFX single crystals. Top: View on the $ab$ plane of the unit cell with four \DMTTF molecules which are rotated by 90$^{\circ}$ with respect to each other. Bottom: Stacks of \DMTTF molecules along the $c$ axis. Each double molecule of such a stack hosts one charge, thus effectively forming a spin chain along the $c$ axis. Crystallographic data are taken from Ref.~\cite{Fourmigue2008}.}
	\label{fig:structure}
\end{figure}

Crystals of \DMTTFCl and \DMTTFBr used in this work were synthesized by electrocrystallization methods as described in Ref.~\cite{Fourmigue2008}. Their structural, electronic and basic magnetic characterization was reported in Refs.~\cite{Fourmigue2008,Foury-Leylekian2011}. The crystallographic structure of the title compounds is shown in Fig.~\ref{fig:structure} and the main features are briefly recapitulated in the following in order to facilitate the discussion of our results. The tetragonal unit cell (space group $I\overline{4}2d$ (no. 122) \cite{Fourmigue2008}) comprises eight planar \DMTTF molecules which form stacks along the crystallographic $c$ axis (Fig.~\ref{fig:structure} bottom) and which are rotated by 90$^{\circ}$ with respect to the neighboring stacks in the $ab$ plane (Fig.~\ref{fig:structure} top). As a consequence of this rotation, interactions between neighboring stacks are weak, leading to effectively one-dimensional electronic properties \cite{Fourmigue2008,Foury-Leylekian2011}. In \DMTTFX the \DMTTF molecules are only partially oxidized and each double molecule within a stack hosts one hole (and three electrons). As a consequence, the electronic bands are quarter filled (three-quarter filled if electrons are considered instead of holes) which gives rise to a metallic behavior at room temperature and at ambient pressure \cite{Fourmigue2008,Foury-Leylekian2011}. Resistivity measurements on \DMTTFX crystals revealed the onset of charge localization below temperatures of $\sim$150\,K \mbox{($X$ = Cl)} and $\sim$100\,K ($X$ = Br), respectively \cite{Foury-Leylekian2011}. Thus, below these temperatures and above the spin-Peierls transition temperatures \TSP $\sim$50\,K of the respective compounds \cite{Foury-Leylekian2011} the spin systems can be described by an $S = 1/2$ Heisenberg AFM chain model in which each double molecule of a stack (see Fig.~\ref{fig:structure} bottom) corresponds to one site in the spin-chain lattice which is occupied by a single spin associated with the localized hole. \DMTTFX crystals grow in a needle-like shape with the longest side of the crystal corresponding to the $c$ axis, and, thus, to the stacking direction of the \DMTTF molecules, which allowed an easy orientation of the crystals in the external magnetic field with respect to this axis. Data presented in this study were collected on several crystals of \DMTTFCl and \DMTTFBr, respectively, in order to avoid artifacts arising from cracks in the crystal caused by thermal cycling. The consistency between measurement series obtained from different crystals was always checked carefully to allow proper comparisons between the different data sets.

Continuous wave (CW) ESR measurements were performed using Bruker EMX X-band spectrometers operating at microwave (mw) frequencies of about 9.6\,GHz and in fields up to 1.3\,T. These spectrometers are equipped with standard He-flow cryostats which allow measurements in the temperature range from 4\,K up to room temperature. Moreover, goniometers installed at the spectrometers enable measurements of ESR parameters as a function of angle between the crystal axes and the external magnetic field. For pulse ESR measurements at X-band frequencies ($\sim$9.7\,GHz) Bruker Elexsys E580 spectrometers were used. These are equipped with a He-flow cryostat and a cryogen-free cryostat, respectively, to provide access to temperatures between 2.7\,K and room temperature.

\section{Results and discussion}
\label{sec:results}

In the following, the results of CW and pulse ESR studies on \DMTTFCl and \DMTTFBr are presented and discussed. For all measurements shown here the external magnetic field was oriented either parallel or perpendicular to the $c$ axis or was rotated in the plane comprising these two directions, respectively. Measurements on \DMTTFBr in the $ab$ plane (not shown) did not reveal a substantial anisotropy of the resonance fields or the linewidth, in line with expectations based on crystal structure (Fig.~\ref{fig:structure} top) and with previous reports \cite{Fourmigue2008,Foury-Leylekian2011}. Consequently, for measurements with $H \perp c$ the crystallographic axes which are parallel to the magnetic field are not specified here since they are magnetically equivalent.

First, we will turn to the results obtained by means of CW ESR measurements over a wide temperature range, Sec.~\ref{subsec:results_CW_ESR}. This will be followed by an investigation of the angular dependence of the CW ESR parameters at various temperatures in the vicinity of and below the spin-Peierls transition, respectively. In Sec.~\ref{subsec:results_pulse_ESR}, a more detailed study of spin dynamics in the low-temperature phase by means of pulse ESR experiments will be presented.

\subsection{CW ESR}
\label{subsec:results_CW_ESR}

\subsubsection{Temperature dependence}

\begin{figure*}
	\centering
	\includegraphics[width=\textwidth]{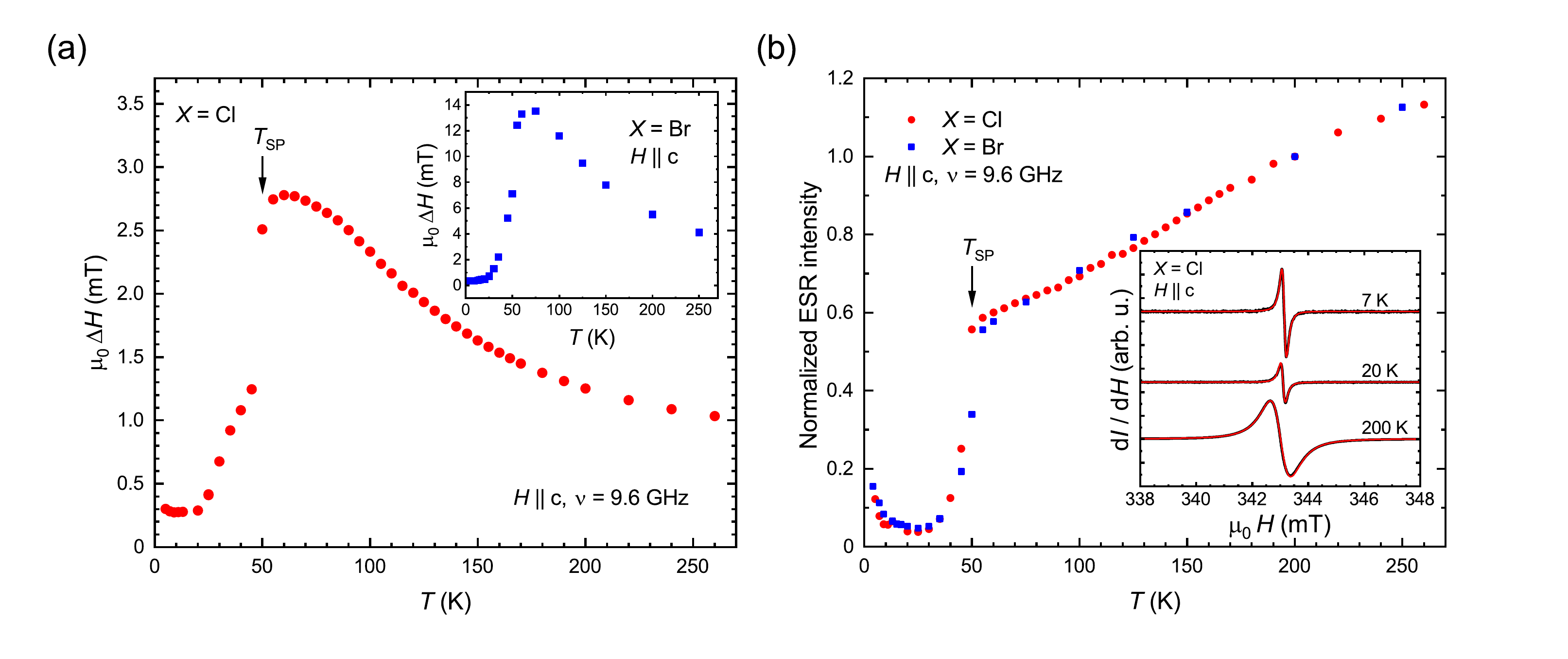}
	\caption{Temperature dependence of the ESR parameters obtained from CW ESR measurements on \DMTTFX samples with the static field applied along the $c$ axis. (a) ESR linewidth  $\Delta H$ of \DMTTFCl. Respective data for \DMTTFBr are shown in the inset. (b) Temperature dependence of the intensity of the ESR line for both compounds obtained from double integration of the spectra. Representative spectra measured on a \DMTTFCl sample at 7, 20, and 200\,K are presented in the inset together with fits to the experimental data (solid red lines). At these different temperature regimes the ESR lines are governed by the magnetic response of the solitons (7\,K), by a combination of soliton and triplon contributions (20\,K), and by the response of the spins within the chains (200\,K), respectively.}
	\label{fig:CW_ESR_T-dep}
\end{figure*}

The temperature dependence of the width as well as the intensity of the ESR line are summarized in Fig.~\ref{fig:CW_ESR_T-dep}. Data shown were measured with the external magnetic field oriented parallel to the $c$ axis, i.e., the spin chain axis. Three representative spectra obtained from measurements on \DMTTFCl at 7, 20, and 200\,K are presented in the inset of Fig.~\ref{fig:CW_ESR_T-dep}(b). As will be explained below, these spectra illustrate the ESR response of the studied system in different temperature regimes which differ with respect to the contributions to the resonance line. Note that, due to the use of the lock-in technique in the CW ESR measurements, the first derivative of the microwave power reflected from the cavity with respect to the external field d$P$/d$H$ is measured. As a consequence, the intensity of the ESR line (i.e., the area under the absorption curve) which is proportional to the static susceptibility $\chi_0$, can be obtained by double integration of the measured spectra. The shown spectra consist of a single resonance line which can be described by a Lorentzian profile. At temperatures between room temperature and above 200\,K, an asymmetric Dysonian lineshape \cite{Feher1955,Dyson1955} was observed which evidences the metallic behavior of the samples in this temperature region \cite{Fourmigue2008,Foury-Leylekian2011}. From fits of Lorentzian or Dysonian lines to the measured spectra [solid red lines in the inset of Fig.~\ref{fig:CW_ESR_T-dep}(b)] the ESR parameters such as linewidth $\Delta H$ (full width at half maximum) and resonance position $H_{\text{res}}$ could be obtained. The latter was used to calculate the (effective) $g$ factor according to the standard frequency-field relation of a paramagnet \cite{AbragamBleaney}
\begin{equation}
\label{eq:resonance_condition}
g = \frac{h\nu}{\mu_{\text{B}}\mu_0 H_{\text{res}}} \ \ ,
\end{equation}
where $\nu$ denotes the mw frequency and $h$, $\mu_{\text{B}}$, and $\mu_0$ are Planck's constant, the Bohr magneton, and the vacuum permeability, respectively. As will be discussed in the following section, angular dependent measurements of the $g$ factor evidence the absence of a substantial temperature dependence of this quantity [Fig.~\ref{fig:CW_ESR_ang-dep}(b)].

The temperature dependence of the linewidth is shown for \DMTTFCl and \DMTTFBr in the main panel and in the inset of Fig.~\ref{fig:CW_ESR_T-dep}(a), respectively. For both compounds, the linewidth reveals a maximum at temperatures slightly above \TSP which is followed towards low temperatures by a sudden drop of the linewidth below \TSP indicating a strong change in spin dynamics below the phase transition. Furthermore, the latter is evidenced also by the steep decrease of the ESR intensity which sets in at \TSP, see Fig~\ref{fig:CW_ESR_T-dep}(b). Down to $\sim$25\,K the intensity, as a measure for the spin susceptibility, is exponentially reduced since the majority of spins approaches the spin-dimerized nonmagnetic singlet ground state expected for a spin-Peierls system. Below $\sim$25\,K the intensity shows a Curie-like increase towards lower temperatures which was previously ascribed to the presence of extrinsic magnetic defects with a spin quantum number $S = 1/2$ \cite{Foury-Leylekian2011}. However, as will be argued in the following sections, the ESR response at low temperatures is, most likely, related to the existence of solitons pinned to naturally present defects in the crystals which show a coherent spin dynamics, similar to the situation found in the related compounds (TMTTF)$_2$AsF$_6$ and (TMTTF)$_2$SbF$_6$ \cite{Bertaina2014}. The different temperature regimes found in the temperature dependences of the ESR linewidth and intensity are, naturally, as well visible in the spectra presented in the inset of Fig.~\ref{fig:CW_ESR_T-dep}(b). At high temperatures (cf. spectrum recorded at 200\,K), the ESR spectrum consists of a resonance line with high intensity and a linewidth of about 1.3\,mT at 200\,K. This line originates from the magnetic response of the spins within the regular \DMTTF-stacks, i.e. the spin chains. Below the spin-Peierls transition, the simultaneous reduction of the linewidth and intensity (caused by the dimerization of the magnetic lattice) yields a rather narrow and weak ESR signal at 20\,K. As will be discussed in the following section, angular dependent measurements reveal that at these temperatures the ESR spectrum, despite featuring only a single line, is composed of two contributions. The dominant contribution to the ESR signal is due to the above-mentioned solitons while the second contribution, arising from singlet-triplet excitations of the dimerized chains, is much weaker and cannot be resolved explicitly in the individual spectra, see below. Upon further lowering the temperature, the latter contribution is strongly reduced and the ESR signal at 7\,K is fully dominated by the solitons whose intensity features a Curie-like behavior, thereby leading to an increased signal intensity as compared to the spectrum at 20\,K.

As mentioned in the Introduction, it is in principle possible to extract the characteristic parameters of the Hamiltonian~(\ref{eq:HAFM-Hamiltonian}), such as $J$ and $\delta$, from an analysis of the temperature dependent spin susceptibility. These parameters, in turn, determine the spin gap $\Delta$ of the system as well as the soliton properties, for instance their spatial extension \cite{Nakano1980,Dobry1997}. Moreover, temperature dependent measurements of the spin susceptibility provide a means for obtaining estimates of the relative amount of solitons with respect to the majority of the spins by comparing the Curie-like tail at low temperatures with the susceptibility at elevated temperatures. Thus, the temperature dependence of the ESR intensity [Fig~\ref{fig:CW_ESR_T-dep}(b)] can yield, in the ideal case, valuable insights into the basic static properties of solitons in a dimerized spin chain system. In the present case, however, such an accurate analysis is hampered by several difficulties as  discussed in more detail in the Appendix where order-of-magnitude estimates for the above-mentioned parameters are presented. Therefore, the temperature dependence of the ESR intensity is discussed on a qualitative level in the following. At temperatures above \TSP, the behavior of the susceptibility resembles the characteristic temperature dependence of uniform $S = 1/2$ Heisenberg AFM spin chains systems \cite{Kluemper1998,Johnston2000,Eggert1994}. On the other hand, a metallic behavior was reported for \DMTTFCl and \DMTTFBr above the charge-localization temperatures of $\sim$150 and $\sim$100\,K, respectively \cite{Foury-Leylekian2011}. This indicates a delocalized character of the charge carriers in this temperature range and, thus, could imply a limited applicability of the model of localized spins. However, below the charge localization temperature, the modeling of \DMTTFX in terms of a spin chain with increasing dimerization below \TSP can be expected to be a valid description. Consequently, the respective model will be employed in the remainder of this paper since the main focus of interest in this work lies on the particular spin dynamics in the spin-gap phase. The existence of such a phase is unambiguously evidenced by the strong decrease of the spin susceptibility below the spin-Peierls transition as mentioned above and allows a study of the dynamic properties of pinned solitons without perturbation by the magnetic response of the spins within the chain. 

Finally, we note that the overall temperature dependence of the linewidth [Fig.~\ref{fig:CW_ESR_T-dep}(a)], in particular, the strong narrowing of the ESR line as well as the temperature-dependent susceptibility [Fig.~\ref{fig:CW_ESR_T-dep}(b)] was observed as well in previous studies \cite{Fourmigue2008,Foury-Leylekian2011} on samples originating from the same source and demonstrates the good agreement between those studies and the measurements of the present work. Thereby it is confirmed that the basic magnetic properties, in particular, the spin-Peierls temperature are the same for samples used in this work and in previous reports. In the following sections, we will extend the discussion of magnetic properties contained in previous studies by examining in more detail the angular dependence of the linewidth at temperatures below \TSP and the spin dynamics at low temperatures.

\subsubsection{Angular dependence}

\begin{figure*}
	\centering
	\includegraphics[width=\textwidth]{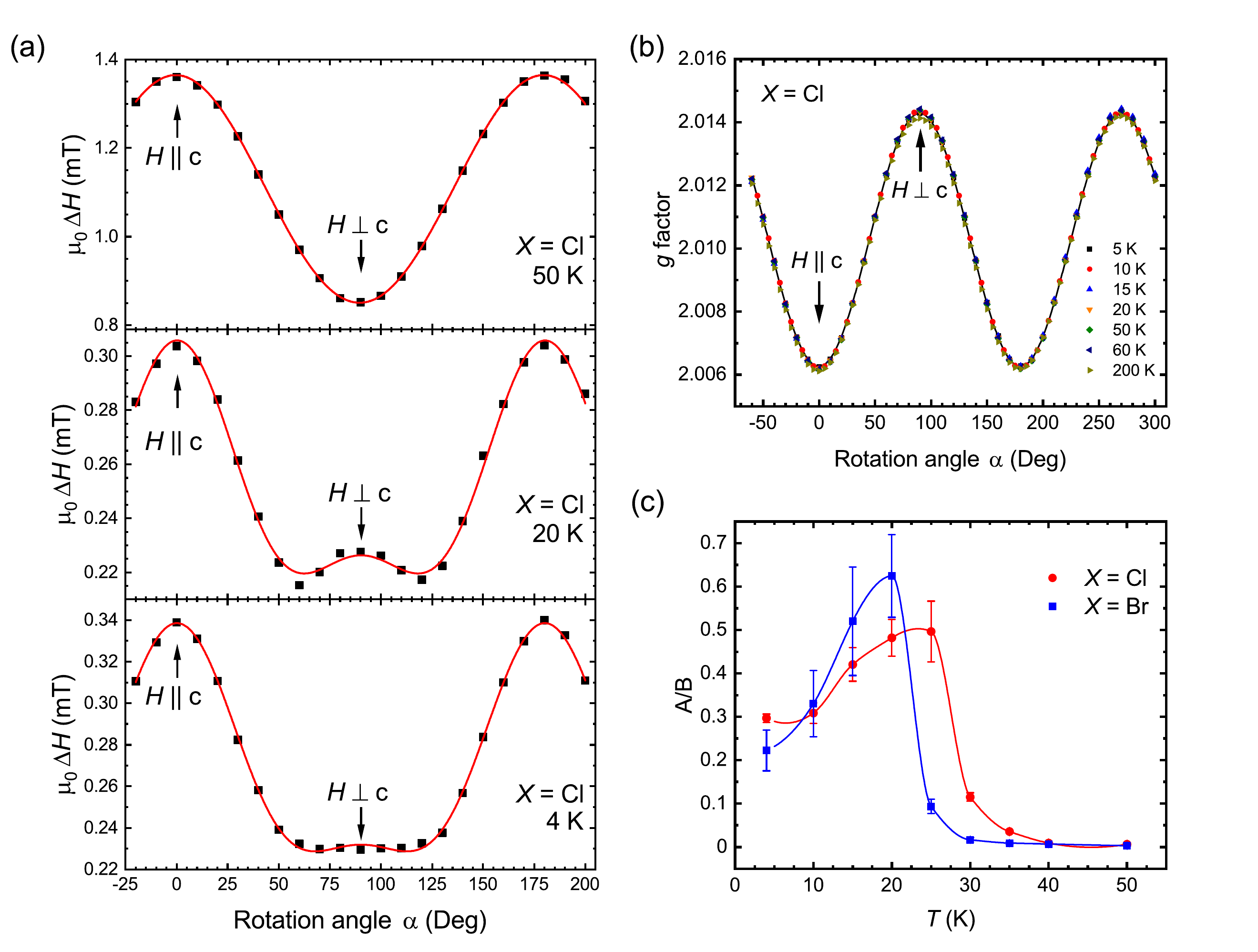}
	\caption{Angular dependence of ESR parameters obtained from CW ESR studies at $\sim$9.6\,GHz. In these measurements the field was rotated from $H \parallel c$ to $H \perp c$. (a) Angular-dependent linewidth of \DMTTFCl at three different temperatures below \TSP showing the effects of different contributions to $\Delta H (\alpha)$ below the transition. Solid lines show fits according to Eq.~(\ref{eq:linewidth_angular_dependence}), see text. (b) Angular dependence of the $g$ factor measured on a \DMTTFCl crystal at temperatures between 5 and 200\,K. Within this temperature range no changes in the $g$ anisotropy are observed. The solid black line denotes a representative fit to the angular dependence of the $g$ factor at 5\,K according to Eq.~(\ref{eq:g-factor_angular_dependence}). Corresponding measurements carried out on \DMTTFBr yielded very similar results as in (a) and (b) and are not shown here. (c) Temperature dependence of the ratio $A/B$ as a measure of the relative strength of the triplon contribution to the linewidth obtained from fits of Eq.~(\ref{eq:linewidth_angular_dependence}) to the $\Delta H (\alpha)$ curves of \DMTTFCl and \DMTTFBr at various temperatures. Solid lines are guides to the eye.}
	\label{fig:CW_ESR_ang-dep}
\end{figure*}

Let us now turn to the discussion of the angular-dependent CW ESR measurements in which the external magnetic field was rotated from $H \parallel c$ (rotation angle $\alpha = 0^{\circ}$) to $H \perp c$ ($\alpha = 90^{\circ}$) at various temperatures. Figure~\ref{fig:CW_ESR_ang-dep}(b) shows the angular dependence of the $g$ factor obtained for \DMTTFCl at temperatures between 5 and 200\,K. Over the whole temperature range under study there is no significant change in the observed $g(\alpha)$ dependence visible, which is an important insight for two reasons. First, since the symmetry of the $g$ tensor is governed by the local surrounding of the resonating spin, the absence of a change in $g(\alpha)$ shows that the local environment remains unchanged even below the spin-Peierls transition which is driven by a lattice dimerization along the $c$ axis of the crystals \cite{Foury-Leylekian2011}. The angular dependence of the $g$ factor can be well described by the following relation for a $g$ tensor with uniaxial symmetry \cite{AbragamBleaney}
\begin{equation}
\label{eq:g-factor_angular_dependence}
g(\alpha) = \sqrt{g_{\parallel}^2\cos^2(\alpha) + g_{\perp}^2\sin^2(\alpha)} \ \ ,
\end{equation}
where $g_{\parallel}$ and $g_{\perp}$ are the principal values of the $g$ tensor corresponding to the external field applied parallel and perpendicular to the $c$ axis, respectively. This type of symmetry is consistent with the shape and the orientation of the \DMTTF molecules within the stacks (see Fig.~\ref{fig:structure}) whose molecular planes are oriented perpendicular to the stacking axis thereby leading to $g_{\perp} > g_{\parallel}$. This is in agreement with the measured $g(\alpha)$ dependence. A representative fit of Eq.~(\ref{eq:g-factor_angular_dependence}) to the data obtained at 5\,K is shown by a black solid line in Fig.~\ref{fig:CW_ESR_ang-dep}(b). It is worthwhile mentioning that very similar results were obtained for \DMTTFBr and are thus not shown here.

Second, the almost identical anisotropies of the $g$ tensor at low and high temperatures clearly evidence that the magnetic response observed over the whole investigated temperature range originates from spins residing on the sites of the spin chain lattice. Therefore, the ESR lines at low-temperatures are not due to magnetic impurities which are unrelated to the spin chains. Instead, these ESR signals can be arguably ascribed to the solitons extending over several lattice sites in the chain and giving rise to a paramagnetic response on the background of the nonmagnetic spin-dimerized state below \TSP (cf. the schematic illustration in Fig.~\ref{fig:spin-Peierls_chain}). 

While the angular dependence of the $g$ factor does not change over the whole temperature range considered here, the angular dependence of the linewidth $\Delta H (\alpha)$ reveals a very pronounced temperature dependence as shown in Fig.~\ref{fig:CW_ESR_ang-dep}(a) for selected temperatures. Above 40\,K the angular dependence of the linewidth shows a $(\cos^2(\alpha) + 1)$ behavior which is regularly encountered in weakly correlated or three-dimensional exchange-narrowed spin systems (see, e.g., Refs.~\cite{Benner1990,Dietz1971} and references therein). Below $\sim$35\,K the angular dependence of the linewidth changes gradually and an additional contribution of the type $(3\cos^2(\alpha)-1)^2$ becomes visible which results in a minimum at $\alpha \approx 60^{\circ}$ and in a local maximum of $\Delta H$ at $\alpha = 90^{\circ}$. Such a contribution could be associated with additional broadening due to unresolved lines originating from $S = 1$ entities (for instance singlet-triplet excitations of the dimerized ground state, denoted as triplons) which are split due to anisotropic couplings (e.g., a dipolar coupling between the triplons) and which are centered symmetrically around the resonance field of the main line \cite{Camara2010,Coulon2004,Bencini2012}. As illustrated by the exemplary spectrum recorded at 20\,K [inset in Fig.~\ref{fig:CW_ESR_T-dep}(b)], this additional contribution to the angular dependence of the linewidth is, indeed, not visible as a separate feature in the spectrum and, thus, can be attributed to an additional unresolved contribution to the resonance line. Lowering the temperature further, finally leads to a complete flattening of the local maximum around $\alpha = 90^{\circ}$ at 4\,K. In order to analyze the described changes in the angular dependence, all $\Delta H (\alpha)$ dependences measured between 4 and 50\,K were fitted with the following empirical expression:
\begin{equation}
\label{eq:linewidth_angular_dependence}
\Delta H (\alpha) = A(3\cos^2(\alpha)-1)^2 + B(\cos^2(\alpha)+1) + \Delta H_0 \ \ .
\end{equation}
The parameters $A$ and $B$ are prefactors which scale the respective contribution to the angular dependence of the linewidth and are thus a measure of the relative strength of these contributions. The offset parameter $\Delta H_0$ describes the isotropic part of the linewidth arising, for instance, from inhomogeneous broadening. From these fits, the temperature evolution of the relative strength of the triplon and chain or soliton (at low temperatures) contributions as expressed in the ratio $A/B$ could be obtained. The resulting temperature dependence of $A/B$ is presented in Fig.~\ref{fig:CW_ESR_ang-dep}(c). Down to 40\,K the ratio is close to zero which confirms the pure $(\cos^2(\alpha)+1)$ character of $\Delta H (\alpha)$ at these intermediate temperatures. Below 40\,K, $A/B$ increases and reaches a maximum at around 25 and 20\,K for \DMTTFCl and \DMTTFBr, respectively. The decrease of $A/B$ below these maximum temperatures is in agreement with a thermally activated behavior anti\-cipated for singlet-triplet excitations and thus supports the scenario of an unresolved $S = 1$ contribution as an origin of the change in $\Delta H (\alpha)$. Within the framework of this approach, one could relate the maximum in $A/B$ to the characteristic energy of the triplons which, in turn, provides a spectroscopic estimate for the energy scale of the singlet-triplet gap of the dimerized chain. The maximum of the triplon intensity at temperatures comparable with the singlet-triplet gap arises from a combination of the thermally activated excitation of triplet states and the Curie-Weiss-like temperature dependence of the excited triplons (see, e.g., \cite{Bencini2012}). This characteristic maximum energy is slightly higher for \DMTTFCl compared to \DMTTFBr. Qualitatively, however, a very similar temperature dependence of $A/B$ was found for both systems indicating that the observed coexistence of several distinct excitations (i.e., triplons and solitons) is a common feature of the investigated compounds. Note that this spin-gap estimate reflects a mean gap value averaged over all temperatures since the spin gap in a spin-Peierls system generally features a pronounced temperature dependence \cite{Johnston2000}. Despite the fact that the spin gap at zero temperature $\Delta(0)$ could be much larger than our spectroscopic estimates (see also the Appendix), the latter nonetheless represent an experimentally accessible, empirical measure of the triplon excitation energies in \mbox{\DMTTFX}. Regarding the temperature evolution of the $A/B$ ratio at the high-temperature side of the maximum, i.e. at temperatures above $\sim$30\,K, we note that the decrease of $A/B$ coincides with the onset of a broadening of the main line [cf. Fig.~\ref{fig:CW_ESR_T-dep}(a)]. Thus, the additional triplon contribution to the ESR signal, as evidenced by a modulation of the angular dependence of the linewidth, could be masked by the overall broadening of the observed resonance line which hampers a detailed analysis of the triplon's temperature dependence at these elevated temperatures.

It is worthwhile mentioning that in ideal one-dimensional spin systems an angular dependence of the linewidth of the form $|3\cos^2(\alpha)-1|^{4/3}$ is expected \cite{Dietz1971} which resembles the observed $(3\cos^2(\alpha)-1)^2$ dependence. However, such a characteristic angular dependence of the linewidth due to the reduced dimensionality should be visible, if present at all, for all temperatures below the charge localization temperatures in \DMTTFX where a spin chain model could be applicable. This is in contrast to the observation of the $(\cos^2(\alpha)+1)$ dependence above $\sim$35\,K, which could be caused, for instance, by residual interchain couplings and consequently rules out the low dimensionality as an origin of the changes in $\Delta H (\alpha)$.

In summary, temperature and angular dependent CW ESR measurements on \DMTTFX crystals showed that the magnetic response in the spin-gap phase below \TSP is governed by the formation of solitons around defect sites or chain breaks and by thermally activated triplons on the nonmagnetic background of the singlet ground state. In the following section the spin dynamics of the pinned solitons, investigated by means of pulse ESR measurements at low temperatures, will be discussed.

\subsection{Pulse ESR}
\label{subsec:results_pulse_ESR}

\begin{figure*}
	\centering
	\includegraphics[width=\textwidth]{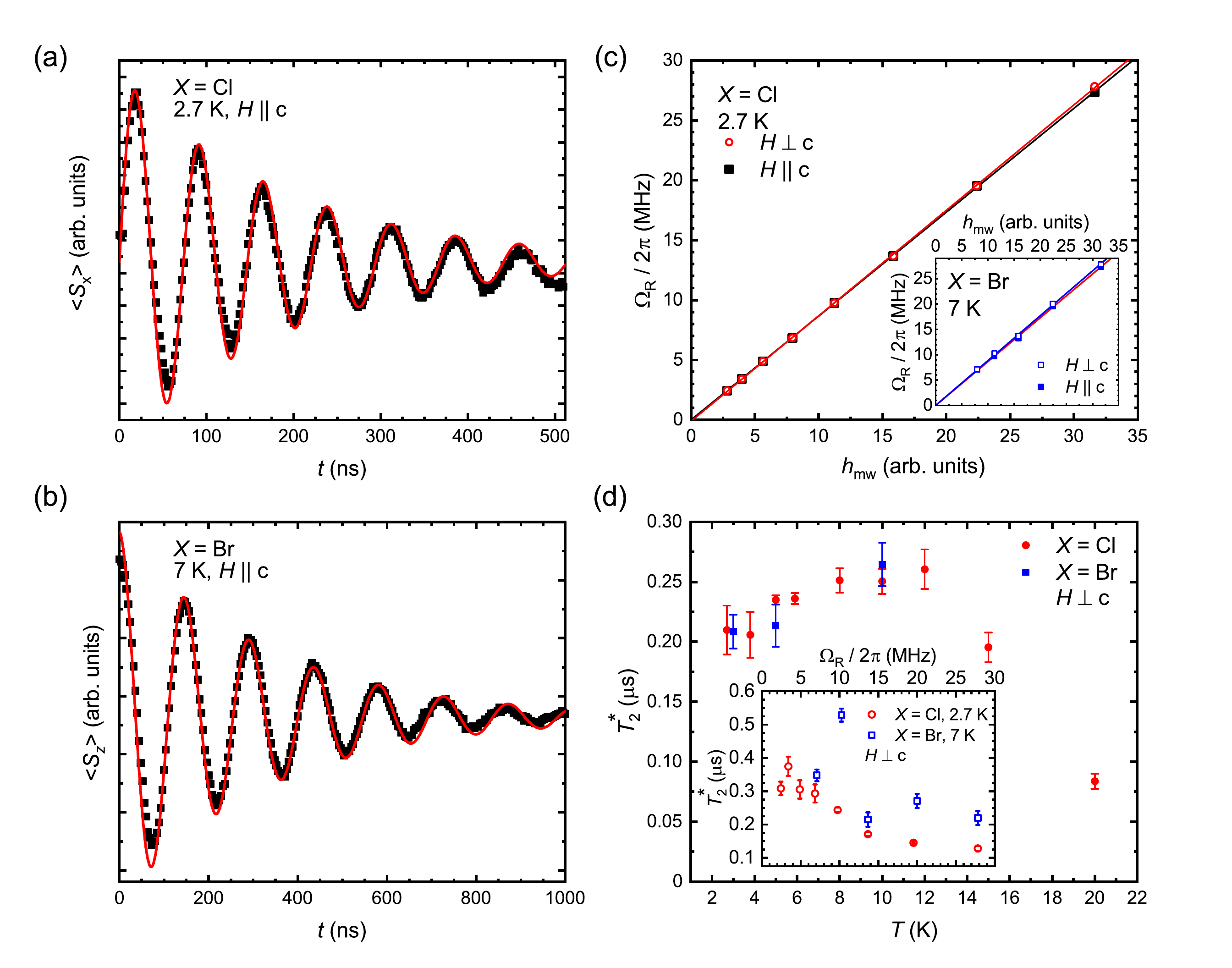}
	\caption{Rabi oscillations in \DMTTFX measured at a mw frequency of $\sim$9.7\,GHz. (a) Time evolution of $\langle S_x (t) \rangle$ in the mw field for \DMTTFCl at 2.7\,K with the external magnetic field applied along the $c$ axis and a mw magnetic field $h_{\text{mw}}$ of about 0.5\,mT. (b) Time evolution of $\langle S_z (t) \rangle$ in the mw field for \DMTTFBr at 7\,K with $H \parallel c$ and $h_{\text{mw}} \sim$0.25\,mT. Solid red lines are simulations of the damped oscillations according to Eq.~(\ref{eq:Rabi_oscillation}). (c) Rabi frequencies $(\Omega_R/2\pi)$ obtained for \DMTTFCl at 2.7\,K and various mw powers, i.e., different $h_{\text{mw}}$. Corresponding data for \DMTTFBr at 7\,K are shown in the inset. Note that the mw magnetic fields are given in arbitrary units since $h_{\text{mw}}$ was not calibrated independently. Solid lines are linear fits to the data. (d) Temperature dependence of $T_2^{\ast}$ determined from the damping of the Rabi oscillations measured on \DMTTFCl and \DMTTFBr crystals with $h_{\text{mw}} \sim$0.4\,mT ($\Omega_R/2\pi \sim$10.5\,MHz) and $h_{\text{mw}} \sim$0.6\,mT ($\Omega_R/2\pi \sim$14.9\,MHz), respectively. The external magnetic field was applied perpendicular to $c$. The inset shows the dependence of $T_2^{\ast}$ as a function of Rabi frequency (and thus $h_{\text{mw}}$) for both compounds measured with $H \perp c$.}
	\label{fig:pulse_ESR_1D_Rabi}
\end{figure*}
In order to shed light on the character of the spin dynamics of pinned solitons in \DMTTFX, nutation measurements were carried out by means of pulse ESR at mw frequencies of about 9.7\,GHz and at low temperatures. In the case of \DMTTFCl a free induction decay (FID) detection was used because of an only weak inhomogeneous broadening of the ESR line. In such a detection scheme a microwave pulse P$_R(t)$ of length $t$ is applied to the sample. The FID signal emitted by the sample and setting in at the end of P$_R$ can be recorded after the deadtime $\tau_d$ ($\sim$80\,ns) of the spectrometer. For recording a Rabi oscillation, the pulse length $t$ is varied and the intensity of the respective FID is measured (see also \cite{DeRaedt2012}). Therefore, the time evolution of the $x$ component of the magnetization $\langle S_x (t) \rangle$ under the influence of a mw magnetic field was measured in these experiments. On the other hand, for \DMTTFBr the ESR line was found to be sufficiently inhomogeneously broadened to allow a detection of the nutation by means of primary spin echoes using the sequence P$_R(t) - T_w - \pi/2 - \tau - \pi - echo$. The waiting time $T_w$ after the end of P$_R$ was chosen to be larger than the phase memory time $T_m$. The time $\tau$ denotes the separation of the two pulses of the standard primary echo sequence. For measuring a Rabi oscillation, the intensity of the spin echo was recorded as a function of $t$. As a consequence, the time evolution of the $z$ component of the magnetization $\langle S_z (t) \rangle$ was measured. Two representative nutation measurements are shown for \DMTTFCl and \DMTTFBr in Figs.~\ref{fig:pulse_ESR_1D_Rabi}(a) and (b), respectively. For both compounds, well-defined Rabi oscillations could be observed which is a clear-cut evidence for a coherent spin dynamics in these compounds at low temperatures. Since the ESR response in this temperature region is dominated by solitons, such a coherent spin dynamics can be attributed to these quantum objects which is similar to the results obtained for other related organic spin chain compounds \cite{Bertaina2014}. As can be seen from Figs.~\ref{fig:pulse_ESR_1D_Rabi}(a) and (b), the Rabi oscillations are damped which is caused by growing decoherence of the soliton spin nutation. The measured Rabi oscillations can be described by the following expressions
\begin{equation}
\label{eq:Rabi_oscillation}
\begin{split}
\langle S_x (t) \rangle &= \langle S_x (0) \rangle\sin(\Omega_Rt)\exp(-t/T_2^{\ast}) \ \ , \\
\langle S_z (t) \rangle &= \langle S_z (0) \rangle\cos(\Omega_Rt)\exp(-t/T_2^{\ast}) \ \ ,
\end{split}
\end{equation}
where $(\Omega_R/2\pi)$ denotes the Rabi frequency of the oscillation and $T_2^{\ast}$ is the empirical damping time to account for decoherence [simulations of the measured oscillations by the latter equations are given as red solid lines in Figs.~\ref{fig:pulse_ESR_1D_Rabi}(a) and (b)]. The two characteristic parameters $(\Omega_R/2\pi)$ and $T_2^{\ast}$ can be obtained from these expressions as well as from the peak in the fast Fourier transform (FFT) of oscillation measurements and a fit of the oscillation envelope by an exponential decay function, respectively. Rabi frequencies of the oscillations as a function of mw magnetic field $h_{\text{mw}}$ at low temperatures are shown for \DMTTFCl in Fig.~\ref{fig:pulse_ESR_1D_Rabi}(c) [corresponding data for \DMTTFBr are presented in the inset]. Note that $h_{\text{mw}}$ is given in arbitrary units because the mw magnetic field at the sample position in the cavity was not calibrated independently to avoid a potential influence of the reference signal (e.g., from a DPPH standard sample) on the Rabi oscillations of the investigated samples. Absolute values given for $h_{\text{mw}}$ in this work are estimates derived from the measured Rabi frequencies and the corresponding $g$ factors determined separately from CW ESR measurements (see Sec. \ref{subsec:results_CW_ESR}). In these calculations it is assumed that $S = 1/2$ for the resonating spins which is justified by the spin-1/2 character of the solitons responsible for the ESR response at low temperatures. As can be seen in Fig.~\ref{fig:pulse_ESR_1D_Rabi}(c), the Rabi frequency shows a linear $h_{\text{mw}}$ dependence which is emphasized by the solid lines that represent linear fits to the data. Such a linear behavior proofs that the Rabi oscillations are indeed measured at the resonance position of the ESR spectrum (i.e., the static external magnetic field and the mw frequency fulfill the resonance condition). Moreover, measurements with the external magnetic field oriented parallel and perpendicular to the $c$ axis revealed only a small anisotropy of the Rabi frequencies, which is caused by the small $g$-factor anisotropy [see Fig.~\ref{fig:CW_ESR_ang-dep}(b)] and the relatively small microwave magnetic field applied [see also Eq.~(\ref{eq:2D_Rabi}) below]. 

The temperature dependence of $T_2^{\ast}$ measured on a \mbox{\DMTTFCl} and a \DMTTFBr sample with \mbox{$H \perp c$} is presented in Fig.~\ref{fig:pulse_ESR_1D_Rabi}(d). Below $\sim$15\,K the relaxation times do not show a significant change upon lowering the temperature. Above 15\,K $T_2^{\ast}$ decreases strongly and above 20\,K the broadening of the ESR line, as observed in the CW ESR measurements [Fig.~\ref{fig:CW_ESR_T-dep}(a)], prevents the detection of an FID or spin echo signal since the broadening is related to faster spin relaxation. Thus, the damping of the Rabi oscillations reflects the behavior of the CW ESR linewidth because both quantities are governed by relaxation processes in the spin system. Compared to (TMTTF)$_2$PF$_6$ \cite{Bertaina2014}, the Rabi oscillations in both \DMTTFX salts are damped more strongly and thus show shorter coherence times. As the decoherence responsible for the damping could be caused by dipolar interactions between the solitons as well as by super-hyperfine interactions between the solitons and the nuclei in their vicinity, a stronger damping would indicate a stronger (anisotropic) interaction of the solitons with their local environment. On the other hand, it was proposed in Ref.~\cite{Bertaina2014} that isotropic exchange interactions between the solitons could stabilize the coherent Rabi oscillations due to exchange averaging of local inhomogeneities caused by dipolar or hyperfine interactions. In this scenario, the exchange between neighboring solitons could be weaker in \DMTTFX for instance due to larger distances between adjacent solitons. Thus it can be concluded that the coherent spin dynamics appears to be a common property of pinned solitons in organic gapped spin chain systems while details of this particular spin dynamics, such as characteristic decoherence or damping parameters, will depend on the specific structural and magnetic properties of the spin chain compounds which determine the (anisotropic) interactions between the solitons as well as their lateral extensions. Moreover, one should note that in addition to the above mentioned contributions, damping could be caused as well by inhomogeneities in the mw magnetic field or local distributions of $g$ tensors \cite{DeRaedt2012,Baibekov2011}. In these cases, 1/$T_2^{\ast}$ is expected to show a linear dependence on the mw magnetic field and thus depends linearly on the Rabi frequency \cite{DeRaedt2012,Baibekov2011}. As a consequence of an inhomogeneous mw magnetic field distribution within the cavity, the relaxation time $T_2^{\ast}$ should decrease with increasing sample size. Therefore, differences between absolute values of $T_2^{\ast}$ between \mbox{\DMTTFX} and \mbox{(TMTTF)$_2X$} \cite{Bertaina2014} could also be related to different sample sizes, if inhomogeneities of $h_{\text{mw}}$ contribute to the damping. In the inset of Fig.~\ref{fig:pulse_ESR_1D_Rabi}(d) the relaxation time $T_2^{\ast}$ as a function of Rabi frequency is shown for \DMTTFCl and \DMTTFBr at 2.7 and 7\, K, respectively. In these measurements, the external static magnetic field was oriented perpendicular to the $c$ axis. Measurements with $H \parallel c$ yielded very similar results and are not shown. For both compounds under study a decrease of $T_2^{\ast}$ with increasing Rabi frequency, i.e., with the strength of $h_{\text{mw}}$ was found which indicates that inhomogeneities of $h_{\text{mw}}$ could indeed affect the measured oscillations (although the data do not follow an ideal linear dependence). 

\begin{figure}[t!]
	\centering
	\includegraphics[width=\columnwidth]{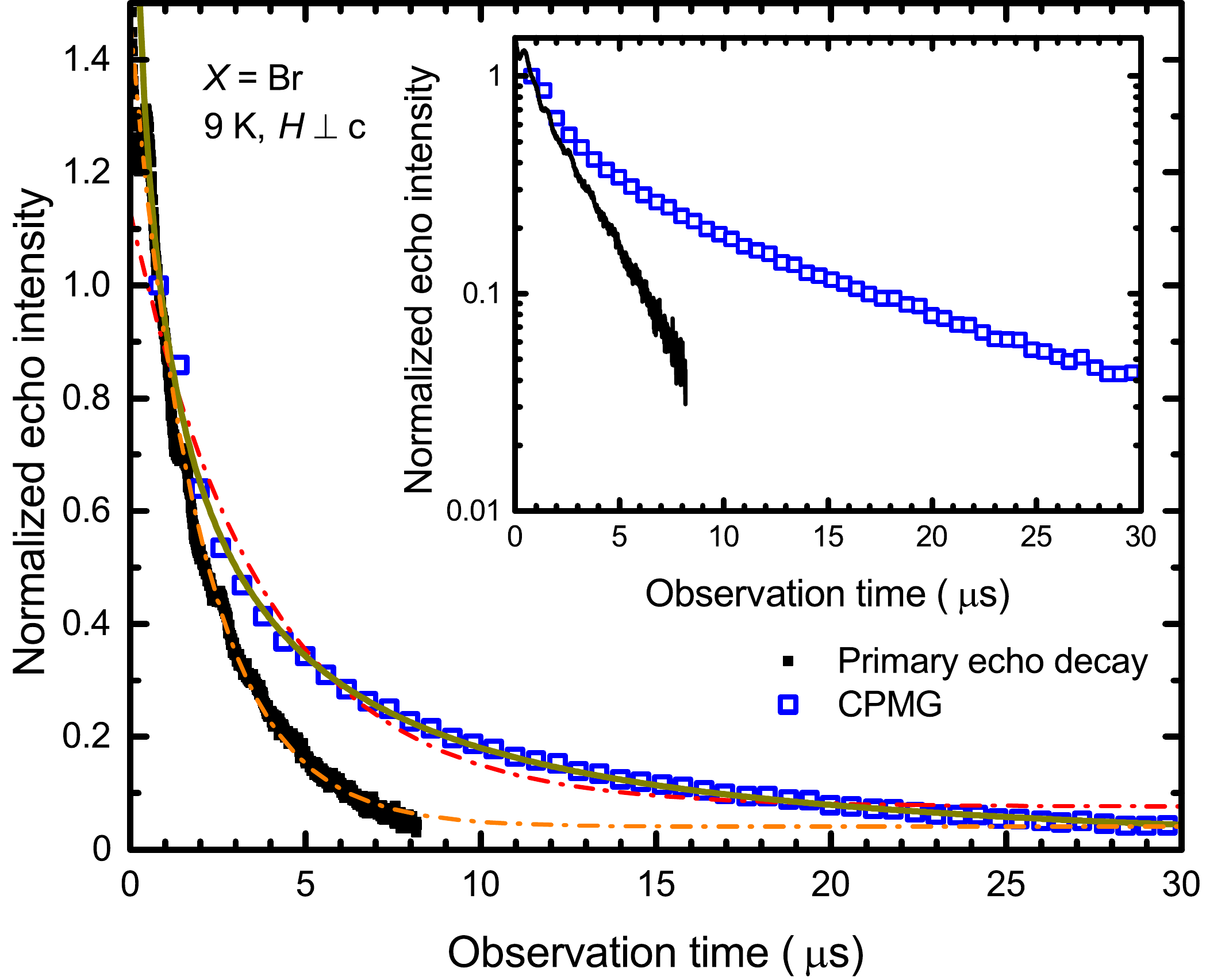}
	\caption{Comparison of primary echo decay (black squares) and CPMG (open blue squares) measurements on \mbox{\DMTTFBr} at 9\,K with $H \perp c$. In both cases a $\pi/2$ pulse of 36\,ns was used. For the CPMG sequence the number of $\pi$ pulses $n$ was set to 50. For comparison, intensities of the spin echoes are normalized to the respective intensities at $t_{\text{ob}} \sim$807\,ns, the time at which the first echo in the CPMG measurement was detected. Dashed-dotted lines denote monoexponential fits to the data while the solid line represents the fit of a stretched exponential function to the CPMG data, see text. The measured data are shown on a semi-logarithmic scale in the inset to emphasize the increase of decoherence times by the CPMG protocol as well as the deviation of the CPMG data from a monoexponential behavior.}
	\label{fig:relaxation_times}
\end{figure}

In order to further investigate the spin dynamics and potential origins of spin decoherence in the Rabi oscillations, the spin relaxation times were measured on \mbox{\DMTTFBr} crystals using pulse ESR techniques. In particular, the phase memory time $T_m$ was measured by means of the primary echo decay method while the transverse relaxation time $T_2$ was studied using the so called Carr-Purcell-Meiboom-Gill (CPMG) protocol \cite{Carr1954,Meiboom1958}. In the former case the pulse sequence \mbox{$(\pi/2)_x - \tau - (\pi)_x - \tau - echo$} was employed and  the intensity of the spin echo was measured as a function of the separation between the pulses $\tau$ (indices of a pulse indicate the direction along which the mw magnetic field is applied during the pulse). For the $T_2$ measurements the sequence $(\pi/2)_x - \left\lbrace \tau - (\pi)_y - \tau - echo - \right\rbrace^n$ was used where $n$ denotes the number of applied $\pi$ pulses. In the CPMG measurements a transient signal is recorded which consists, in the ideal case, of $n$ echoes. Analyzing the decrease of the echoes' heights then allows a determination of the (effective) $T_2$ (see below). Examples of such measurements which were conducted at 9\,K with the external static magnetic field applied perpendicular to the $c$ axis are shown in Fig.~\ref{fig:relaxation_times}. In these measurements a $\pi/2$-pulse length of 36\,ns was used together with 16-step and 2-step phase cycling in the primary echo decay and CPMG measurements, respectively. For the CPMG sequence the number of $\pi$ pulses $n$ was set to 50. For comparison of the relaxation behavior obtained with the two different techniques, the measured spin echo intensities were normalized to the respective intensities at $t_{\text{ob}} \sim$807\,ns, i.e., the time at which the first echo was detected in the CPMG measurement. In this representation of the data it becomes evident that, qualitatively, the relaxation in the spin system is faster when a simple primary echo decay method is employed while the CPMG protocol yields a slower relaxation. Such an enhancement has been theoretically predicted and experimentally observed, for instance, in molecular magnets (see, e.g., Refs.~\cite{Zaripov2013,Zaripov2017} and references therein) and was attributed to contributions of certain spectral diffusion processes, caused by a stochastic modulation of the hyperfine interactions, to the decoherence of the spins which are strongly suppressed by the CPMG pulse sequence. In this case, however, the shape of the relaxation curves of the spin echoes is expected to be converted from a stretched-exponential behavior in the case of the primary echo method towards a monoexponential decay in the case of CPMG measurements \cite{Zaripov2017}. In the present study, an opposite behavior was found: While the primary echo decay curves could be satisfactorily described by a monoexponential decay, the corresponding CPMG measurements clearly deviate from this behavior. In Fig.~\ref{fig:relaxation_times} fits to the data according to the following expression are shown
\begin{equation}
I(t_{\text{ob}}) = I(0)\exp(-(t_{\text{ob}}/T_{2,m})^b) + I_{\text{off}} \ \ \ .
\label{eq:stretched_exp_decay}
\end{equation}
Here, $I(t_{\text{ob}})$ represents the echo intensity at the observation time $t_{\text{ob}}$, $T_{2,m}$ denotes the respective relaxation time ($T_m$ for primary echo decay and $T_2$ for the CPMG method, respectively), $b$ is the stretching parameter, and $I_{\text{off}}$ is an empirical constant offset to describe the measured data. The dashed-dotted lines in Fig.~\ref{fig:relaxation_times} represent fits with $b$ set to 1, i.e., monoexponential fits, which yield $T_m \sim$2.0\,$\mu$s for the primary echo decay and $T_2 \sim$3.7\,$\mu$s for the CPMG measurements. As mentioned above, in the latter case the monoexponential fit does not yield a proper description of the data. The quality of the fit can be significantly improved by treating $b$ as an independent fit parameter which results in a strongly reduced effective relaxation time of about 0.5\,$\mu$s and $b \sim$0.36 (the corresponding fit is shown as solid line in Fig.~\ref{fig:relaxation_times}). In particular the small $b$ value indicates a spread of relaxation times which inhibits a meaningful quantitative comparison of the relaxation times obtained by employing the two different pulse sequences. Note that fitting a stretched exponential to the data of the primary echo measurements yields a very similar value for $T_m$ and $b \sim$0.96, which further evidences the monoexponential nature of the primary echo decay. In addition, the consequences of the different applied pulse protocols for the character of the spin decoherence are emphasized in the inset of Fig.~\ref{fig:relaxation_times} which presents the measured data on a semi-logarithmic scale. In this representation, the data of the primary echo decay measurements follow a straight line, as expected for a monoexponential decay, while the CPMG data show a non-linear behavior but overall a slower relaxation of the spins as compared to the primary echo method. As pointed out, for instance, in Ref.~\cite{Zaripov2013} the CPMG method is well-suited to suppress the spin decoherence due to spectral diffusion originating from hyperfine interactions. Despite the qualitative differences of the relaxation curves observed in the present study and in Ref.~\cite{Zaripov2013}, the relevance of hyperfine interactions for the understanding of spin dynamics in \mbox{\DMTTFX} is evidenced by the appearance of a modulation of the primary echo decay curves (see Fig.~\ref{fig:relaxation_times}), the so-called \mbox{ESEEM} effect \cite{Mims1972,Schweiger2001} which is caused by dipolar \mbox{(super-) hyperfine} interactions between electron spins and the nuclei in their vicinity. Thus, it appears to be very instructive to study details of the hyperfine interactions between the solitons and their environment by means of more advanced, highly sensitive techniques such as \mbox{HYSCORE} spectroscopy \cite{Schweiger2001} which is, however, beyond the scope of the present study. It is worthwhile mentioning that, within the temperature range of our CPMG investigation between 3.6 and 16\,K, the strength of deviations of the CPMG decay curves from a monoexponential behavior increase with decreasing temperature. This could point towards a temperature-dependence of the spin decoherence mechanism, for instance by virtue of temperature dependent spin-lattice relaxation of the (electron or nuclear) spins \cite{Zaripov2013,Zaripov2017}. In conclusion, it is possible to effectively enhance the spin coherence of the solitons in \DMTTFBr by applying a CPMG pulse sequence. However, this pulse protocol induces at the same time a distribution of relaxation times which hampers a simple characterization of the spin coherence in terms of a single $T_2$. Whether this peculiar behavior is an artifact of temperature dependent spin-lattice relaxation processes or indeed an intrinsic property of the soliton spin dynamics remains an open question requiring further studies. Nonetheless, we would like to emphasize that the phase memory times $T_m$ of the order of 2\,$\mu$s determined in the measurements discussed here are in any case larger than the damping times $T_2^{\ast}$ of the Rabi oscillations which did not exceed $\sim$0.55\,$\mu$s [see Fig.~\ref{fig:pulse_ESR_1D_Rabi}(d)]. Therefore, spectral diffusion, although potentially active in \DMTTFBr as evidenced by the difference between primary echo and CPMG decay curves, appears to be not the main source of decoherence in the observed Rabi oscillations. Instead, the relaxation of the soliton spins in the nutation experiments might be dominated by inhomogeneities of the mw magnetic field as was proposed in the discussion of the $h_{\text{mw}}$ dependence of $T_2^{\ast}$ (see above).

\begin{figure*}[t!]
	\centering
	\includegraphics[width=\textwidth]{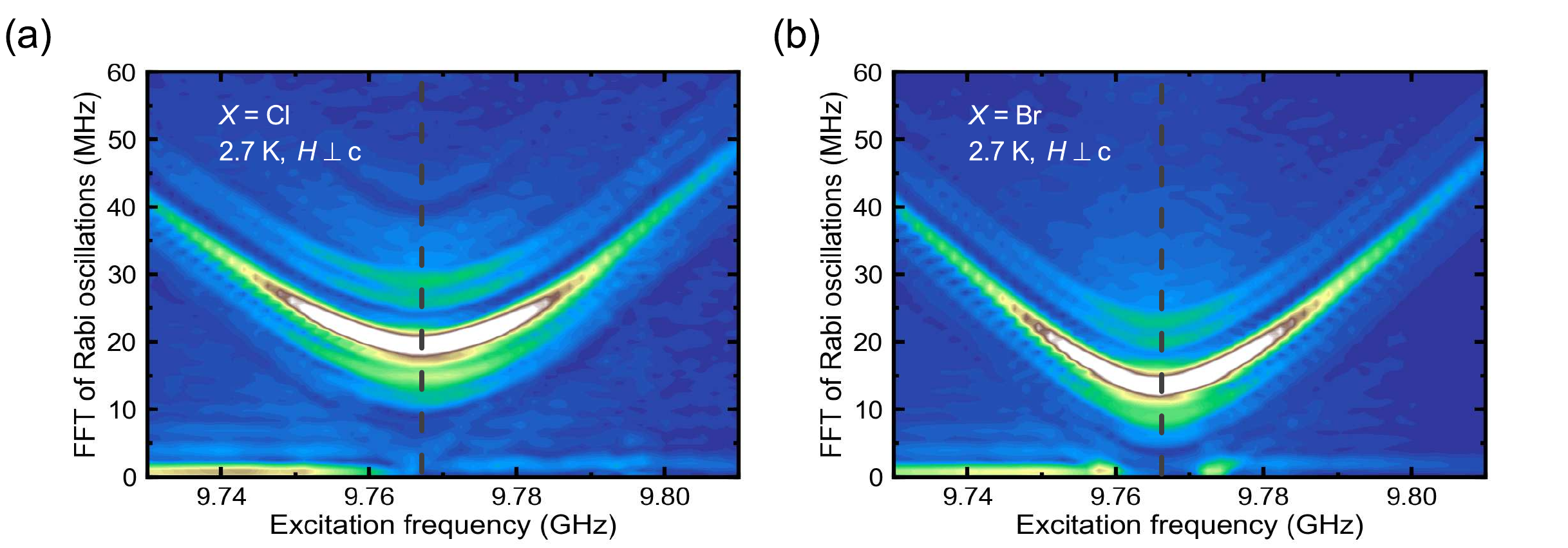}
	\caption{Fourier transform of Rabi oscillations with different excitation frequencies at 2.7\,K and with $H \perp c$ for (a) \DMTTFCl and (b) \DMTTFBr, respectively. The intensity of the fast Fourier transform (FFT) is given as color code where blue corresponds to low and light colors represent high intensities. The external magnetic field was set to match a resonance frequency of about 9.765\,GHz which is indicated by gray dashed lines. These measurements were performed with microwave powers which correspond to microwave magnetic fields of about 0.7 and 0.6\,mT for \DMTTFCl and \DMTTFBr, respectively.}
	\label{fig:pulse_ESR_2D_Rabi}
\end{figure*}

In addition to the temperature and the $h_{\text{mw}}$ dependence of the nutation measurements presented in the preceding paragraphs, Rabi oscillations were measured using different excitation frequencies centered around the resonance frequency in order to search for the existence of additional contributions to the coherent oscillations, for instance, from $S = 1$ triplet excitations. The FFTs of the oscillations obtained from such detuning experiments at 2.7\,K with $H \perp c$ are shown for \DMTTFCl and \DMTTFBr in Figs.~\ref{fig:pulse_ESR_2D_Rabi}(a) and (b), respectively. In this representation the intensity of the FFT (in arbitrary units) is color coded where blue corresponds to low intensities while light colors represent regions of high FFT intensity. For both investigated salts very similar FFT spectra were obtained. In particular, the main frequency of the Rabi oscillation is seen as a branch of high intensity which reveals a square-root-like dependence on the detuning parameter $\Delta = h (\nu_{\text{res}} - \nu_{\text{mw}})$. The frequencies $\nu_{\text{res}}$ and $\nu_{\text{mw}}$ are the resonance frequency of the considered transition and the mw frequency used for exciting the Rabi oscillations, respectively. Thus, the observed main branch follows well the relation expected for the so-called Rabi precession (see, e.g., Ref.~\cite{Cohen-Tannoudji2005})
\begin{equation}
\Omega_R /2\pi = \sqrt{\Delta^2 + (g\mu_B h_{\text{mw}})^2}/h \ \ \ ,
\label{eq:2D_Rabi}
\end{equation}
which describes the frequency of a transition between states in a two-level system (for instance, a spin-1/2 system in an external static magnetic field) driven by an external electromagnetic field. The second branches of weaker intensity which are visible for both compounds above the main branches and show the same curvature as the respective main branch are due to an instrumental artifact and are not related to the intrinsic pulse ESR response of the \DMTTFX crystals. Therefore, these measurements at different excitation frequencies showed no evidence of further spin contributions aside from the $S = 1/2$ oscillations of the solitons. 

Despite the differences regarding details of the soliton spin dynamics in organic spin systems discussed here and in previous studies \cite{Bertaina2014}, the existence of Rabi oscillations in these compounds is remarkable, in particular, if compared to other inorganic systems which show similar coherent oscillations \cite{Bertaina2007,Nellutla2007,Rakhmatullin2009,DeRaedt2012,Baibekov2011,Baibekov2017}. For observing the coherent spin dynamics in the latter systems, usually a dilution of magnetic ions is required in order to reduce anisotropic interactions with the environment of the magnetic ions and to reduce interactions between the magnetic ions \cite{DeRaedt2012}. This emphasizes the important role played by the isotropic exchange between the solitons (as pointed out in a previous study \cite{Bertaina2014}) and by the fact that solitons are built of a highly correlated ensemble of entangled spins. Both mentioned aspects appear to be crucial to protect the Rabi oscillations from fast decoherence and thus render them observable in experiments.

\section{Conclusions}
\label{sec:Conclusions}
In conclusion, we presented a CW and pulse ESR study of \DMTTFCl and \DMTTFBr single crystals. These compounds belong to the family of organic gapped spin chain systems. Angular and temperature dependent CW ESR measurements revealed that the magnetic low-temperature properties are governed by solitons pinned to defect sites in the (magnetic) lattice as well as by thermally activated triplons on the nonmagnetic singlet background of the dimerized spin chain. For both compounds very similar characteristic temperatures of a maximal triplon contribution were found which indicates that their spin-gaps are similar as well. The singlet ground state of the majority of spins in the chains allowed a separate investigation of the soliton spin dynamics at low temperatures by means of pulse ESR experiments. Nutation measurements revealed the existence of well-defined Rabi oscillations which is evidence for a coherent character of the spin dynamics. Studies of the spin relaxation times in the system by means of standard primary-echo decay and advanced CPMG methods revealed an enhancement of the coherence times as compared to the characteristic damping times of the Rabi oscillations which shows that the decoherence of the latter is influenced by inhomogeneities in local magnetic fields, such as the mw magnetic field. Although increasing the spin coherence time as compared to the primary-echo decay measurements, the application of the CPMG protocol induced a spread of spin relaxation times in contrast to the behavior expected for spin systems affected by specific types of spectral diffusion \cite{Zaripov2013}. Thus, our work should motivate further detailed investigations of the decoherence mechanisms relevant in \DMTTFX, for instance the (super-) hyperfine interactions between the solitons and their environment. Taken together with similar results obtained on other organic spin chain compounds (Ref.~\cite{Bertaina2014}) it appears that the coherence of the spin dynamics is a specific feature of pinned solitons in gapped spin chain systems with naturally occurring defects, potentially caused by isotropic exchange interactions between the solitons. As a consequence, it is possible to coherently manipulate the $|m_z\rangle$ states of these quantum objects which might be of interest for potential application in quantum computation.

\section*{Acknowledgments}
J.Z. acknowledges financial support from the German Academic Exchange Service (DAAD). ESR measurements at IM2NP and BIP Marseille as well as at LASIR Lille were supported by the CNRS research infrastructure RENARD (IR-RPE CNRS 3443).

\appendix

\section*{Appendix: Analysis of the spin susceptibility}
\label{sec:app_susc}
In this section we analyze the temperature-dependent ESR intensity measured on a \mbox{\DMTTFCl} crystal [Fig.~\ref{fig:CW_ESR_T-dep}(b)] in order to obtain estimates for the parameters of the dimerized chain model which is used to approximately describe the spin system in \DMTTFX. As mentioned in Sec.~\ref{subsec:results_CW_ESR}, an accurate analysis of the $\chi(T)$ data obtained from ESR measurements is hampered by several aspects which will be discussed in the following before presenting the results of the approach employed.

First, a precise evaluation would require the availability of absolute values of the magnetic susceptibility in order to enable consistency checks of the obtained fitting parameters. These tests are needed to connect the experimental data with the theoretical predictions, such as offsets to account for diamagnetic or van Vleck paramagnetic contributions or proportionality factors. Although the ESR intensity shown in Fig.~\ref{fig:CW_ESR_T-dep}(b) is proportional to the spin susceptibility, absolute values cannot be obtained with the used setup. This situation is regularly encountered in ESR spectroscopy and could be overcome by comparison and scaling to data from static magnetic susceptibility measurements if the latter is available (see, for instance, Refs.~\cite{Dumm2000,Deisenhofer2006}). In the case of \DMTTFX only powder measurements of the static susceptibility were reported \cite{Foury-Leylekian2011} due to the weak paramagnetic response of the spins in combination with a strong diamagnetic background which required an increase of sample mass used for the measurements as well as a background correction. Thus, the calibration of ESR intensities from single-crystal measurements against corrected powder measurements of the static susceptibility would introduce uncertainties to the absolute values and was not applied in the present case. Instead, ESR intensities were normalized to the respective 260\,K value and consequently did not allow an evaluation of the spin susceptibility's absolute value.

\begin{figure*}[t]
	\includegraphics[width=\textwidth]{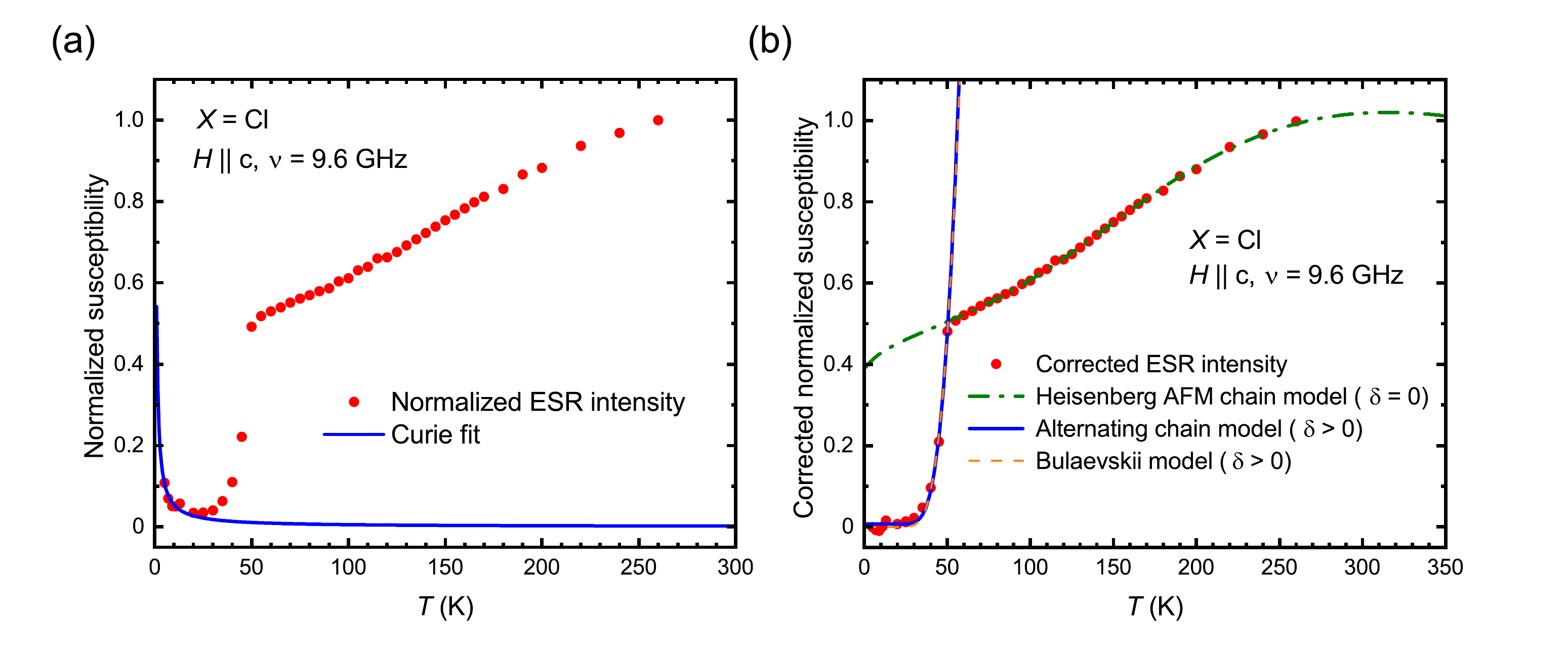}
	\caption{Analysis of the temperature-dependent normalized susceptibility as obtained from CW ESR measurements at 9.6\,GHz on a \mbox{\DMTTFCl} crystal with external field applied parallel to $c$. (a) Fit of a Curie law (blue solid line) to the low-temperature part of the ESR intensity to correct the data with respect to the soliton contribution. (b) Corrected data after subtraction of the fit shown in (a) from the data. Dashed-dotted line denotes a fit for $T > 50$\,K using the model of a Heisenberg AFM chain ($\delta = 0$) from Ref.~\cite{Johnston2000} in order to determine the exchange coupling constant $J$. The fit in the low-temperature region for $\delta > 0$ and fixed $J$ is shown by the blue solid line. For comparison a fit within the same temperature range using the Bulaevskii model \cite{Bulaevskii1969} is indicated by the orange dashed line.}
	\label{fig:temp_dep_susceptibility}
\end{figure*}

Second, when comparing experimental data with theoretical curves one should take into account that measurements are in most cases carried out at constant pressure $p$ while calculations are usually done for systems of constant volumes $V$ \cite{Dumm2000}. If thermal expansion is strong, as it is the case in organic systems, the difference between the magnetic susceptibility measured at constant pressure $\chi^{p = \text{const.}}$ can significantly differ from the susceptibility at constant volume $\chi^{V = \text{const.}}$ which requires corrections based on further measurements \cite{Dumm2000}. Thus the qualitative behavior of the temperature-dependent susceptibility at elevated temperature might be altered due to this effect. In the present case it cannot be decided unambiguously whether the non-linear increase of ESR intensity with increasing temperature is indeed due to the 1D AFM-spin-chain character of the spin system or due to the effect of $\chi^{p = \text{const.}}$ (a transformation from $\chi^{p = \text{const.}}$ to $\chi^{V = \text{const.}}$ was shown to result in a flattening of the $\chi (T)$ curve \cite{Dumm2000}).

This aspect is related to the third difficulty encountered when analyzing the ESR intensity of \mbox{\DMTTFCl}: the proper choice of a model to describe the data. Despite the  metallic behavior at temperatures above $\sim$150\,K \cite{Foury-Leylekian2011} the temperature dependence of the ESR intensity shown in Figs.~\ref{fig:CW_ESR_T-dep}(b) and \ref{fig:temp_dep_susceptibility} resembles qualitatively the one found for other low-dimensional spin systems with a singlet ground state which could be described by an isotropic Heisenberg AFM chain above their respective transition temperatures \cite{Johnston2000,Jacobs1976,Lohmann1997,Law2011}. Although the characteristic maximum in $\chi (T)$ is not completely captured in the temperature range up to 260\,K used for the measurements discussed here, a non-linear increase is nonetheless visible, in particular an inflection point (at around 105\,K) which is typical for one-dimensional AFM spin chains \cite{Eggert1994,Kluemper1998,Johnston2000}. However, one should note that the shape of the $\chi (T)$ curve could be influenced by thermal expansion of the sample (see previous paragraph). The flattening of the temperature dependence of the ESR intensity after a correction for this effect was observed in other organic spin chain compounds \cite{Dumm2000} and could, in principle, lead to an almost temperature-independent spin susceptibility which would be expected for an ideal metal (see, e.g., Ref.~\cite{Blundell2001}). The situation in the systems under study could also be in between these two limiting cases of completely localized charges and delocalized metallic behavior due to an interplay of the hopping between the sites $t$ and the on-site Coulomb repulsion $U$. For particular values of the ratio $t/U$ the susceptibility was modeled by Seitz and Klein \cite{Seitz1974}. In particular, the characteristic maximum in $\chi (T)$ was found to be shifted towards smaller temperatures (in units of $J$) with increasing $t/U$. Thus, the strength of the nearest neighbor exchange coupling derived from the temperature of the maximal ESR intensity depends on the specific value of $t/U$. 

Since these parameters are currently not known for \mbox{\DMTTFX} we decided to analyze the ESR intensity of \DMTTFCl as a first approximation in terms of a chain of localized spins with a uniform coupling above and an alternating coupling below \TSP as described by the Hamiltonian given in the Introduction [Eq.~(\ref{eq:HAFM-Hamiltonian})]. This analysis was carried out along the lines given in Ref.~\cite{Johnston2000}. In a first step we approximated the low-temperature part ($T < 20$\,K) of the normalized intensity curve by fitting a Curie-type law,
\begin{equation}
\chi_{\text{C}} (T) = \frac{C_0}{T} \ \ \ ,
\end{equation}
to the data, in order to describe the contributions of the solitons to the spin susceptibility using the scaling factor $C_0$ [see Fig.~\ref{fig:temp_dep_susceptibility}(a)]. The resulting $\chi_{\text{C}} (T)$ was subsequently subtracted from the measured data to obtain the contribution of the spin chain for further analysis [red circles in Fig.~\ref{fig:temp_dep_susceptibility}(b)]. The temperature range above \TSP was then fitted using the model of an $S = 1/2$ AFM uniform Heisenberg chain ($\delta = 0$) \cite{Johnston2000} to determine $J$ above the spin-Peierls transition by
\begin{equation}
\chi_{\text{chain}} (T) = \chi_0 + \frac{C_{\text{chain}}}{J(1+\delta)}\chi^{\ast}(\delta,T) \ \ \ ,
\label{eq:susceptibility_fit}
\end{equation}
where $\chi^{\ast}$ is the theoretical spin susceptibility of the alternating chain according to Eq.~(56a) in Ref.~\cite{Johnston2000}. The latter equation is the specific form of the general expression Eq.~(50a) in Ref.~\cite{Johnston2000} and is the relevant equation for the case of an alternating chain with $\delta$ as a parameter. Setting $\delta$ to zero recovers the susceptibility of the uniform Heisenberg chain \cite{Johnston2000}. The corresponding fit yielded a $J$ of about 500\,K and is shown in Fig.~\ref{fig:temp_dep_susceptibility}(b) by a dashed-dotted line where the temperature range was extended to 350\,K to visualize the maximum in $\chi (T)$. Considering the fit we note that there are two difficulties when evaluating the reliability of the obtained $J$. First, a negative offset $\chi_0$ of about -1.3 was necessary to describe the data by the model which seems to be unphysical but could be caused by the normalization of the ESR intensities due to the lack of absolute values. And second, the absence of precise quantitative data prevents consistency checks of the obtained proportionality factor $C_{\text{chain}}$ (not only the position but also the height of the maximum in $\chi (T)$ is defined by $J$). Thus, we emphasize that the $J$ and $\delta$ values presented in this section are merely fitting parameters of the simplified model necessarily applied in the analysis. Interestingly, our value of $J$ is of the same order of magnitude as the exchange constants determined for other organic spin chain compounds such as (TMTTF)$_2$PF$_6$ \cite{Dumm2000}. Thus, $J \sim$500\,K appears to be a reasonable value for the isotropic exchange coupling in \DMTTFCl above \TSP. By comparing the values obtained for $C_0$ and $C_{\text{chain}}$ which are proportional to the number of solitons and spins in the chains, respectively, the concentration of the solitons $c_{\text{soliton}} = N_{\text{soliton}}/N_{\text{spin}}$ could be estimated to be 0.0003 using a $g$ factor of 2.006 for this orientation of the sample and $S = 1/2$. Note, however, that the obtained $C_{\text{chain}}$ might be influenced by the offset parameter $\chi_0$. Indeed, a comparison of the value of the corrected normalized ESR intensity [Fig.~\ref{fig:temp_dep_susceptibility}(b)] at the inflection point at around 105\,K with the theoretical curve for the uniform chain ($\delta = 0$) \cite{Kluemper1998} with $J \sim500$\,K yields a smaller amount of spins in the chain and, thus, a value of the soliton concentration of about 0.0009. Although being larger than the concentration obtained directly from the fit parameters, the concentration determined by this method is still of the same order of magnitude as the previous estimate.

Within the framework of the approximations contained in the analysis of the high-temperature part of the spin susceptibility, we described the ESR intensity below \TSP by a model of an alternating spin chain to obtain an estimate for the dimerization parameter $\delta$ according to Ref.~\cite{Johnston2000}. For fitting the ESR intensity below 50\,K, we kept the $J$ obtained in the previous step constant and used $\delta$, $\chi_0$, and $C_{\text{chain}}$ in Eq.~(\ref{eq:susceptibility_fit}) as fit parameters. As a result we obtained $\delta \sim$0.25 but also values for $\chi_0$ ($\sim$0), and $C_{\text{chain}}$ which differ from the fit above \TSP [the fit is shown as solid blue line in Fig.~\ref{fig:temp_dep_susceptibility}(b)]. Moreover, the obtained $\delta$ rather constitutes an estimate for the dimerization parameter at very low temperatures ($T <<$ \TSP) because $\delta$ as well as $J$ can show a temperature dependence due to the gradual development of the lattice dimerization at the spin-Peierls transition \cite{Johnston2000}. Therefore, keeping $J$ constant during the fitting procedure could influence the obtained $\delta$. Furthermore, we compared $\delta$ from the fit according to Johnston \etal \cite{Johnston2000} with the model by  Bulaevskii \cite{Bulaevskii1969} which has been used in previous studies to estimate the dimerization within the chains \cite{Dumm2000,Lohmann1997}. From the corresponding fit, which is shown as dashed orange line in Fig.~\ref{fig:temp_dep_susceptibility}(b), we could derive a $\delta$ of 0.25 in good agreement with the previous value. This analysis relies, however, on the same value of $J$ determined from the high-temperature part of $\chi(T)$ and thus serves only as a self-consistency check of the employed fitting procedure. From the obtained estimates of $J$ and $\delta$ it is possible to calculate the spin gap of the alternating spin chains. Utilizing the expression $\Delta = 2J\delta^{3/4}$ according to Barnes \etal \cite{Barnes1999} yields $\Delta \sim$350\,K, which is an estimate for the zero-temperature gap since the obtained $\delta$ value represents the zero-temperature limit of the dimerization parameter. Note that the derived value of $\Delta$ is much larger than the spectroscopic estimate of the order of 25\,K (see Sec.~\ref{subsec:results_CW_ESR}) which might be caused either by the pronounced temperature dependence of the spin gap (see, for instance, Ref.~\cite{Johnston2000}) or the uncertainties in the determination of $J$ and $\delta$, respectively. 

Due to the above-mentioned approximations used and the limitations caused by the normalization of the ESR intensity one should, faute de mieux, consider the absolute values obtained for $J$, $\delta$, $\Delta$, and $c_{\text{soliton}}$ as order-of-magnitude estimates only. Our experimental work should therefore motivate further theoretical studies of the electronic properties of the considered \mbox{\DMTTFX} salts, in particular calculations of $t$ and $U$, in order to validate the applicability of the utilized model Eq.~(\ref{eq:HAFM-Hamiltonian}) in the presented analysis.

\bibliography{literature_DMTTF}

\end{document}